\RequirePackage{fix-cm}
\documentclass[twocolumn,referee]{svjour3}
\smartqed
\usepackage[T1]{fontenc}
\usepackage{textcomp}
\usepackage{mathptmx}
\usepackage{amsmath}
\usepackage{amsfonts}
\usepackage{amssymb}
\usepackage{makeidx}
\usepackage{graphicx}
\usepackage{color}
\usepackage[numbers,sort&compress]{natbib}
\usepackage{subfigure}
\usepackage[version=3]{mhchem}
\usepackage[colorlinks=true,bookmarksnumbered=true]{hyperref}
\hypersetup{pdftitle={Structure and elastic properties of eta carbide
from ab initio calculations. Application to cryogenically treated gear steel.},
pdfauthor={Adrian Oila},pdfsubject={DFT},pdfkeywords={Fe2C, ab initio},
linkcolor=black,citecolor=black}

\vbadness=\maxdimen

\journalname{Journal of Materials Science}
\begin{document}\sloppy
\title{Elastic properties of eta carbide ($\eta$-Fe$_2$C)
 from \textit{ab initio} calculations. Application to cryogenically treated
 gear steel.}
\author{Adrian Oila \and Chi Lung \and Steve Bull}

\institute{A. Oila \at
              School of Chemical Engineering and Advanced Materials \\
              Newcastle University, Newcastle upon Tyne, NE1 7RU \\
              United Kingdom\\
              \email{Adrian.Oila@newcastle.ac.uk}
}

\date{}
\maketitle
\begin{abstract}
The elastic properties of $\eta$-Fe$_2$C (eta carbide) have been determined
from \textit{ab initio} density functional theory (DFT) calculations using
the generalized gradient approximation (GGA).
The isotropic polycrystalline elastic modulus of $\eta$-Fe$_2$C has been
calculated as the average of anisotropic single-crystal elastic constants
determined from the \textit{ab initio} simulations.
The calculated polycrystalline elastic modulus was used to compute
the elastic modulus of a case carburised gear steel subjected
to shallow cryogenic treatment (SCT) and deep cryogenic treatment (DCT).
This value was then compared with experimental values obtained from
nanoindentation.
The results confirmed that the changes in elastic modulus observed in
the DCT steel can be attributed to the precipitation
of $\eta$-Fe$_2$C. No changes in hardness have been observed
between the SCT steel and the DCT steel.
These data were then used to assess the surface contact fatigue behaviour
of the SCT and DCT steels tested under elastohydrodynamic lubrication
(EHL) conditions.
 
\keywords{Ab initio \and Eta carbide \and Cryogenic \and Contact fatigue}
\PACS{31.15.A \and 61.50.Ah \and 61.66.Fn \and 81.40.Pq \and 81.40.Ef}

\end{abstract}
\section*{Introduction}
\label{sec:introduction}

$\eta$-Fe$_2$C (eta carbide) and $\varepsilon$-Fe$_{2.4}$C (epsilon carbide)
are two transition compounds which occur in the microstructure of quenched
steels during the initial stages of tempering \cite{Nakamura1986}.
The precipitation of $\varepsilon$-Fe$_{2.4}$C is predominant in conventional
heat treatments (quenching in oil at temperatures above $273\,\text{K}$)
while $\eta$-Fe$_2$C precipitates during cryogenic (sub-zero) treatments,
known as \textit{shallow} when the quenching temperature is
near $193\,\text{K}$ and \textit{deep} when the quenching is performed
at or near $77\,\text{K}$ \cite{Baldissera2008,Bensely2006}.

The application of cryogenic treatments to steel components such as
tools \cite{Das2009,Das2010a,Das2010,Firouzdor2008,Molinari2001}
and gears \cite{Baldissera2008,Bensely2006,Baldissera2009,Baldissera2009a,
Bensely2007,Bensely2008,Bensely2009,Bensely2011,
Manoj2004,Paulin1993,Preciado2006,Stratton2009}
is justified by numerous claims that the wear and fatigue behaviour
is significantly improved mainly due to three phenomena which occur
at low temperatures \cite{Baldissera2009a}: complete martensitic
transformation, changes in the residual stresses and precipitation
of nanometric carbides.

The microstructure of surface hardened steels commonly used
to manufacture heavy-duty gears typically consists
of tempered martensite, retained austenite and iron carbides.
The complexity of this microstructure has lead to somewhat
contradictory opinions regarding the role played by individual phases
in wear and contact fatigue. An example for this is the influence of
retained austenite and its optimum amount (a brief review can
be found in \cite{PhD-Oila2003}).

A better understanding of the role played by individual phases
is necessary for reliable failure predictions and this requires
that the mechanical properties of the phases involved are known.
Experimental determination of these properties (i.e. elastic modulus,
hardness, yield strength, etc.) can be difficult, on one hand because
of the small size of the grains (the $\eta$-Fe$_2$C observed \cite{Meng1994}
varies from 5 to 10 nm in cross-section and from 20 to 40 nm in length) and,
on the other hand because some phases are not stable at room temperature
(i.e., unalloyed Fe-C austenite). The structure of Fe-C austenite
as well as a number of relevant properties have been computed
by molecular dynamics \cite{Oila2009} but, to date no experimental
or theoretical data exists for the elastic modulus of $\eta$-Fe$_2$C.
 
The lattice parameter of $\eta$-Fe$_2$C has been determined from
\textit{ab initio} calculations by various authors
\cite{Ande2012,Bao2009,Faraoun2006,Lv2008}, its bulk modulus has
also been calculated \cite{Faraoun2006,Lv2008} but the anisotropic
single-crystal elastic constants have been computed only by
Lv \textit{et al.} \cite{Lv2008}.

Although, the mechanisms by which cryogenic treatments improve the wear
resistance of steels are not completely understood it is believed
\cite{Meng1994} that the precipitation of nanometric $\eta$-Fe$_2$C
enhances the strength and toughness of the martensite matrix,
similar to the reinforcement of composites with nanoparticles.
Also, the precipitation of the nanometric carbides is accompanied
by a reduction in residual stresses in martensite \cite{Preciado2006}.
The proposed mechanism \cite{Meng1994} of $\eta$-Fe$_2$C formation
at low temperatures involves a slight shift of carbon atoms
from the equlibrium position due to lattice deformation.

In this work, we determined the structural and elastic properties
of $\eta$-Fe$_2$C from first principles. These include the lattice
parameters and the single-crystal elastic constants.
The isotropic polycrystalline elastic moduli have been calculated
as averages of single-crystal elastic constants using the Hill's
average \cite{Hill1952}.

The calculated elastic modulus for $\eta$-Fe$_2$C and the experimentally
determined elastic modulus of martensite were used to estimate
the elastic modulus of a gear steel subjected to two different
cryogenic treatments: shallow (SCT) and deep (DCT), respectively
by applying the rule of mixtures (Eq. \ref{eq:13}).
These data were then used to assess the contact fatigue behaviour
of the steel tested under rolling/sliding elastohydrodynamic
lubrication (EHL) conditions.
 
At the time of writing there is no published work on the wear behaviour of
cryogenically treated gear steels under EHL conditions (in which most
case carburised gears operate). 
\section*{Ab initio calculations}
\label{sec:abinitio}
\begin{center}
\begin{figure}[h!]
  \includegraphics[width=0.5\textwidth]{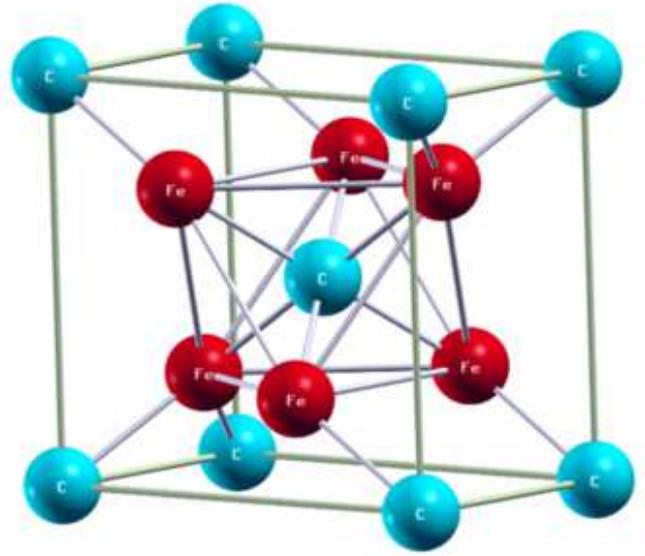}
  \caption{The orthorhombic unit cell of $\eta$-Fe$_2$C.}
  \label{fig:Fig1}
\end{figure}
\end{center}

The crystal structure of $\eta$-Fe$_2$C (Fig. \ref{fig:Fig1})
is orthorhombic \cite{Dirand1983,Hirotsu1972,Nagakura1981,Nakamura1986},
space group $Pnnm$ (58), with 6 atoms in the conventional
unit cell: 4 Fe atoms and 2 C atoms. The Wyckoff positions
of the atoms are Fe $4g$ $(x,0.25,0)$ and C $2a$ $(0,0,0)$.
The experimentally measured lattice parameters \cite{Dirand1983}
$a=4.704$ \AA, $b=4.318$ \AA{} and $c=2.830$ \AA{} were used as initial values
in the simulations.
The \textit{ab initio} spin-polarized calculations were performed employing
the generalized gradient approximation (GGA) \cite{Perdew1996}
as implemented in the Quantum-ESPRESSO package \cite{Giannozzi2009},
using atomic ultrasoft pseudopotentials \cite{Laasonen1993} within the
density functional theory (DFT) \cite{Hohenberg1964,Kohn1965}.
The use of generalized gradient approximation (GGA) is preferred because
it correctly predicts the ferromagnetic body centred cubic (BCC)
structure of Fe, while the local density approximation (LDA) incorrectly
predicts its ground state to be nonmagnetic \cite{Jiang2008}.
 
\begin{center}
\begin{figure}[h!]
  \includegraphics[width=0.5\textwidth]{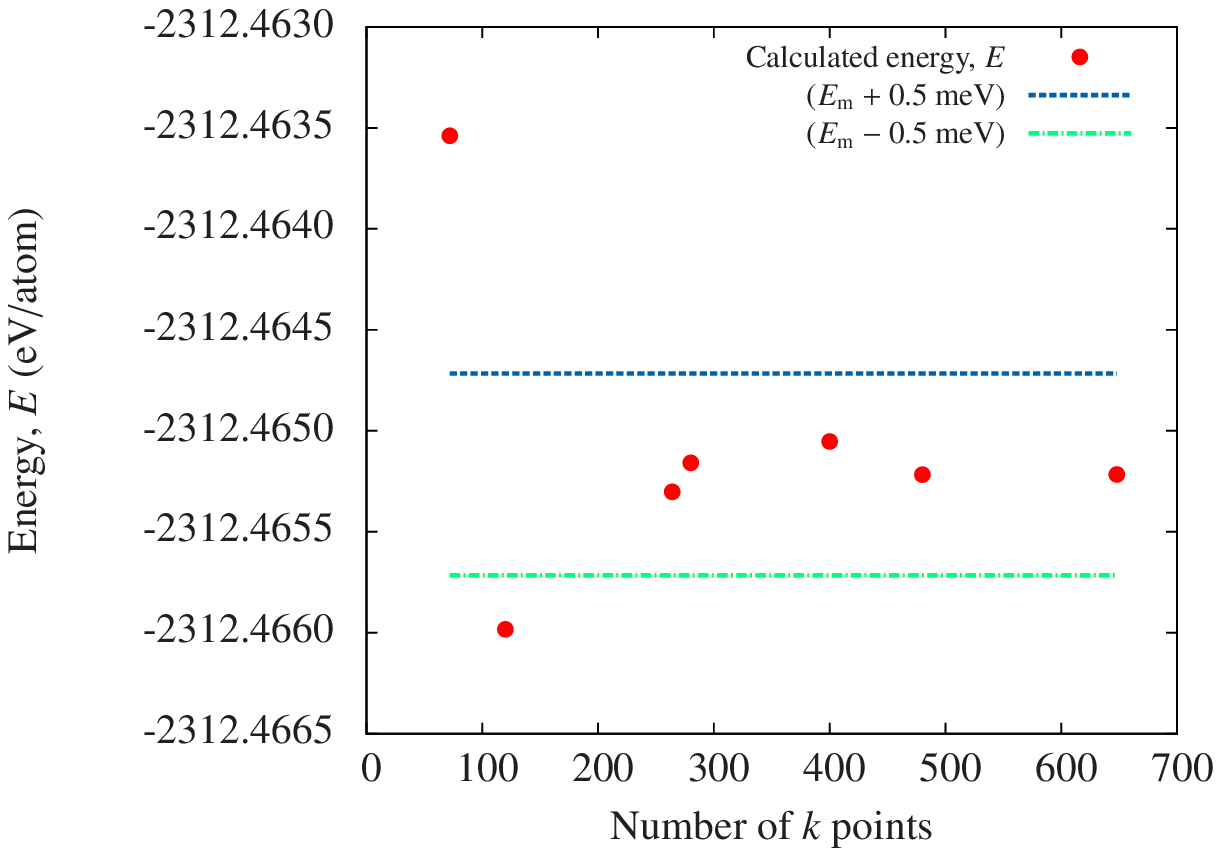}
  \caption{Energy function of the number of $k$-points in the Brillouin zone.
   $E_m = \text{the energy computed using the most dense mesh.}$}
  \label{fig:Fig2}
\end{figure}
\end{center}

The Brillouin zone was sampled by constructing a $k$-points mesh
following the Monkhorst-Pack scheme \cite{Monkhorst1976}
in which the $k$-points are homogeneously distributed in rows
and columns running parallel to the reciprocal vectors.
The Brillouin zone integrations were performed using
a Marzari-Vanderbilt method \cite{Marzari1999} with a
Gaussian spreading of 0.005 Ry ($\sim 0.068$ eV).
A mesh $6 \times 7 \times 10$ which gives 264 $k$-points in the Brillouin zone
was selected for consequent calculations.
Fig. \ref{fig:Fig2} shows the energy values computed for
different $k$-points meshes. For meshes containing 264 or more
$k$-points all energy values lie within a window of $1 \, \text{meV}$.
 
\begin{center}
\begin{figure}[h!]
  \includegraphics[width=0.5\textwidth]{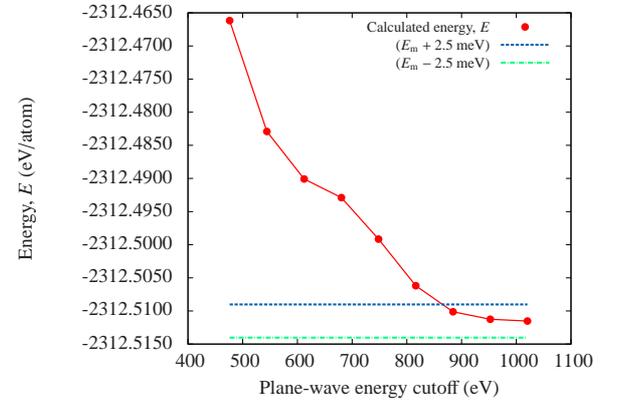}
  \caption{Energy convergence test for $\eta$-Fe$_2$C.
   $E_m = \text{the energy computed using the highest energy cutoff.}$}
  \label{fig:Fig3}
\end{figure}
\end{center}

After the convergence tests, a plane wave kinetic-energy
cutoff of 65 Ry ($\sim 884$ eV) and a charge density cutoff of 390 Ry ($\sim 5306$ eV)
were found to be sufficient to converge the total energy
to less than 5 meV/atom (Fig. \ref{fig:Fig3}).

\begin{center}
\begin{figure}[h!]
  \includegraphics[width=0.5\textwidth]{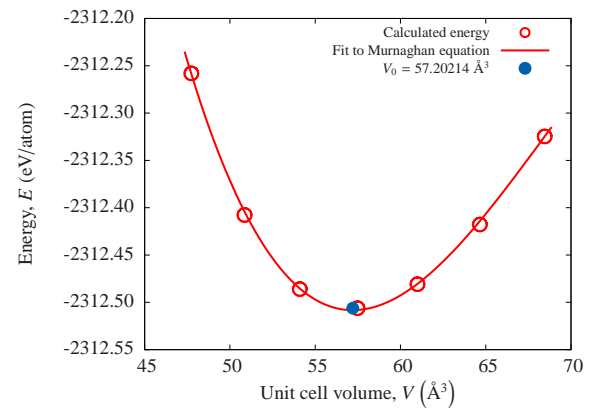}
  \caption{Calculated energy as a function of the $\eta$-Fe$_2$C unit cell volume.}
  \label{fig:Fig4}
\end{figure}
\end{center}

Structural optimisation was performed by computing the total
energy as a function of the unit cell volume by varying the 
$b/a$ and $c/a$ ratios while allowing the atomic coordinates to relax
according to a conjugate-gradient scheme.
The calculated energy was plotted versus the unit cell volume (Fig. \ref{fig:Fig4})
and fitted to the Murnaghan equation of state \cite{Murnaghan1944}.

\begin{figure}[h!]
  \includegraphics[width=0.45\textwidth]{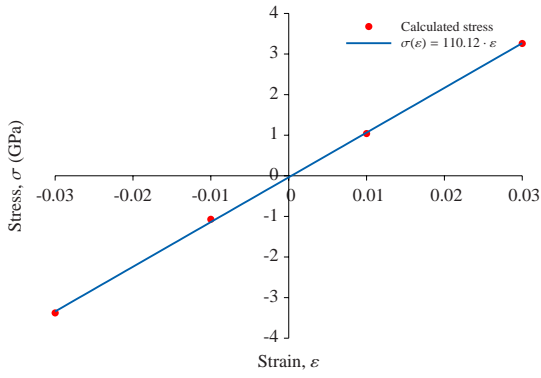}
  \caption{Calculation of elastic constant $c_{44}=110.12 \, \text{GPa}$.}
  \label{fig:Fig5}
\end{figure}

The elastic constants, $c_{ij}$, of the orthorhombic unit cell
were calculated by applying a small strain to the equilibrium
lattice parameter and computing the total energy.
The symmetric distortion matrix for an orthorhombic unit cell, 
$\vec{D}$, is given by \cite{Jiang2008}:

\begin{equation}
 \vec{D} = \left( \begin{array}{ccc}
1 + \varepsilon_1 & \varepsilon_6 / 2 & \varepsilon_5 / 2 \\
\varepsilon_6 / 2 & 1 + \varepsilon_2 & \varepsilon_4 / 2 \\
\varepsilon_5 / 2 & \varepsilon_4 / 2 & 1 + \varepsilon_3 \end{array} \right)
\label{eq:1}
\end{equation}

where $\varepsilon_i$ are the strain tensor components in Voigt
notation. The elastic constants, $c_{ij}$, can be calculated from the 
Hook's law. Fig. \ref{fig:Fig5} shows an example of linear fitting of the
calculated stress versus the applied strain. The corresponding elastic constant
represents the slope of the fitted curve.

The polycrystalline bulk modulus $B$ (Eq. \ref{eq:2}) and shear modulus $G$ (Eq. \ref{eq:3})
can be calculated using the Hill's average \cite{Hill1952}:

\begin{eqnarray}
 B = \dfrac{B_{Reuss} + B_{Voigt}}{2}
\label{eq:2}
\end{eqnarray}

\begin{eqnarray}
 G = \dfrac{G_{Reuss} + G_{Voigt}}{2}
\label{eq:3}
\end{eqnarray}

where $B_{Reuss}$ and $G_{Reuss}$ are given by Eq. \ref{eq:4} and \ref{eq:5}
assuming uniform stress \cite{Reuss1929} and $B_{Voigt}$ and $G_{Voigt}$ are given by
Eq. \ref{eq:6} and \ref{eq:7} assuming uniform strain \cite{Voigt1887}.

\begin{eqnarray}
 B_{Reuss} = \dfrac{1}{s_{11} + s_{22} + s_{33} + 2\left(s_{12} + s_{23} + s_{13}\right)}
\label{eq:4}
\end{eqnarray}

\begin{eqnarray}
 G_{Reuss} = \dfrac{15}{4\left(s_{11} + s_{22} + s_{33} - s_{12} - s_{23} - s_{13}\right)
 + 3\left(s_{44} + s_{55} + s_{66}\right)} \nonumber \\
\label{eq:5}
\end{eqnarray}

\begin{eqnarray}
 B_{Voigt} = \dfrac{c_{11} + c_{22} + c_{33} + 2 \left(c_{12} + c_{23} + c_{13}\right)}{9}
\label{eq:6}
\end{eqnarray}

\begin{eqnarray}
 G_{Voigt} = \dfrac{c_{11} + c_{22} + c_{33} - c_{12} - c_{23} - c_{13}}{15} +
 \dfrac{c_{44} + c_{55} + c_{66}}{5} \nonumber \\
\label{eq:7}
\end{eqnarray}

\begin{table*}[ht!]
\caption{Chemical composition of S156 steel, wt \%.}
\begin{center}
\begin{tabular}{cccccccc}
\hline
C & Si & Mn & P & S & Cr & Mo & Ni \\
\hline
0.14-0.18 & 0.10-0.35 & 0.25-0.55 & max. 0.015 & max. 0.012 & 1.00-1.40 & 0.20-0.30 & 3.80-4.30 \\
\hline
\end{tabular}
\end{center}
\label{tab:1}
\end{table*}

The elastic modulus, $E$, and the Poisson's ratio, $\nu$, can be calculated
using Eq. \ref{eq:8} and \ref{eq:9}, respectively.

\begin{eqnarray}
 E = \dfrac{9BG}{3B + G}
\label{eq:8}
\end{eqnarray}

\begin{eqnarray}
 \nu = \dfrac{3B/2 - G}{3B + G}
\label{eq:9}
\end{eqnarray}

\section*{Experimental}
\label{sec:experimental}
Tests were carried out on samples of S156 steel which had been
carburised, quenched and surface ground.
The chemical composition of the S156 steel is given in Table \ref{tab:1}.
The cryogenic treatments (SCT and DCT) were carried out at Frozen Solid UK
after tempering at $190\,^{\circ}$C. 
The depth of the hardened case after grinding was approximately 1 mm.
The surface finish measured by optical profilometry was $R_a=0.2-0.4$ $\mu$m,
a value similar to that commonly obtained in gears.
The retained austenite content has been measured by X-Ray diffraction
using a XSTRESS 3000 (Stresstech Group) stress analyser.
The values corresponding to the depth below surface at which
the nanoindentation tests were carried out ($500$ $\mu$m)
are given in Table \ref{tab:5}.

\subsection*{Nanoindentation}
\label{subsec:nanoindentation}
The nanoindentation tests were carried out using a Hysitron Triboindenter
with a Berkovich tip using a maximum applied load of $10 \, \text{mN}$.
After each indentation an area $5 \times 5 \, \mu\text{m}$ was scanned using
the AFM (Atomic Force Microscope) of the triboindenter.
Hardness, $H$ (Eq. \ref{eq:10}) and elastic modulus, $E$ (Eq. \ref{eq:11} and \ref{eq:12})
have been calculated using the Oliver-Pharr method \cite{Oliver1992}.

\begin{equation}
 H = \dfrac{P_{max}}{A_c}
\label{eq:10}
\end{equation}
\begin{tabbing}
 where:\= \\
 \> $P_{max}$ is the maximum indentation load \\
 \> $A_c$ is the projected area of tip-sample contact
\end{tabbing}

\begin{equation}
 E^* = \dfrac{1}{2} \sqrt{\dfrac{\pi}{A_c}}S
\label{eq:11}
\end{equation}
\begin{tabbing}
 where:\= \\
 \> $E^*$ is the reduced contact modulus \\
 \> $S$ is the stiffness
\end{tabbing}

\begin{equation}
 \dfrac{1}{E^*} = \dfrac{1-\nu^2}{E} + \dfrac{1-\nu_i^2}{E_i}
\label{eq:12}
\end{equation}
\begin{tabbing}
 where:\= \\
 \> $\nu$ and $\nu_i$ are the Poisson's ratios of sample\\
 \> and indenter, respectively \\
 \> $E$ and $E_i$ are the Young's moduli of sample\\
 \> and indenter, respectively
\end{tabbing}

A total of 50 indentations have been performed on a polished
cross section of each sample at a depth of approximately $500$ $\mu$m.
\subsection*{Surface contact fatigue}
\label{subsec:fatigue}
The surface contact fatigue tests have been carried out using a rig 
described in a previous publication \cite{Oila2005b}.
A number of six pairs of discs have been tested: two oil quenched,
two shallow cryogenic treated and two deep cryogenic treated.
In the conventional treatment the samples were oil quenched from $825\,^{\circ}$C,
and tempered at $190\,^{\circ}$C.

In order to achieve an elliptical contact, one of the discs was crowned
with a crown height of $5 \, \text{mm}$, giving a crown radius of $250 \, \text{mm}$.
All contact fatigue tests have been carried out for $5 \times 10^5 \, \text{cycles}$
under a contact pressure $1.5 \, \text{GPa}$, at a temperature of $60 \, ^{\circ} \text{C}$
and a speed of $1200 \, \text{rev/min}$ with a slide-to-roll ratio of 0.33.
The lubricant used was Valvoline HP Gear Oil 85W-140 1/5 GA
and the calculated $\lambda$ ratio varied between 0.2 and 0.5. 

The type of failure on all specimens, as observed by reflected light
microscopy (RLM) and scanning electron microscopy (SEM) was predominantly
micropitting (see Fig. \ref{fig:Fig9}).
The distribution of the micropits inside the contact area, around the circumference
of the disc, follows a uniform pattern which allows for the computation
of the average percentage area of damage.
The percentage damage has been measured by processing the images captured
with a reflected light microscope, and it was determined as
the average of 10 measurements taken in different locations
chosen at random on the disc surface.
\section*{Results and discussion}
\label{sec:results}

\subsection*{Calculated properties of $\eta$-Fe$_2$C}
\label{subsec:structure}
The results calculated in this work have been compared
with those obtained by others (where available).
The calculated lattice parameters are generally in good agreement with
values reported by other authors (Table \ref{tab:2}).
\begin{table}[h!]
\caption{Lattice parameters (\AA) of $\eta$-Fe$_2$C.}
\begin{center}
\begin{tabular}{cccc}
\hline
Source & $a$ & $b$ & $c$ \\
\hline
This study & 4.722 & 4.271 & 2.835 \\
\cite{Hirotsu1972} & 4.704 & 4.318 & 2.830 \\
\cite{Faraoun2006} & 4.687 & 4.261 & 2.830 \\
\cite{Lv2008} & 4.677 & 4.293 & 2.814 \\
\cite{Bao2009} & 4.651 & 4.258 & 2.805 \\
\cite{Ande2012} & 4.708 & 4.281 & 2.824 \\
\hline
\end{tabular}
\end{center}
\label{tab:2}
\end{table}

\begin{table*}[ht!]
\caption{Elastic constants of $\eta$-Fe$_2$C (GPa).}
\begin{center}
\begin{tabular}{cccccccccc}
\hline
Source & $c_{11}$ & $c_{22}$ & $c_{33}$ & $c_{12}$ & $c_{23}$ & $c_{13}$ & $c_{44}$ & $c_{55}$ & $c_{66}$ \\
\hline
This study & 323 & 340 & 378 & 189 & 158 & 136 & 110 & 97 & 136 \\
\cite{Lv2008} & 310 & 346 & 296 & 170 & 216 & 170 & 64 & 148 & 157 \\
\hline
\end{tabular} \\
\end{center}
\label{tab:3}
\end{table*}

The single-crystal elastic constants of $\eta$-Fe$_2$C are presented in Tabel \ref{tab:3}.
There are significant differences between the values calculated
in the present study and those obtained by Lv \textit{et al.} \cite{Lv2008}.
The accuracy of the calculated elastic constants is strongly dependant on the accuracy
of the self consistency runs and also on the convergence criteria of geometry optimizations
for each distorted structure. In our calculations we have used a denser $k$-points mesh
($6 \times 7 \times 10$ compared to $6 \times 6 \times 9$ in \cite{Lv2008}) and
we imposed a convergence threshold of $10^{-8}$ Ry ($\sim 1.36 \times 10^{-7}$ eV) while
the convergence threshold used in \cite{Lv2008} was $10^{-5}$ eV.

The bulk modulus, $B$, (Tabel \ref{tab:4}) calculated in this work agrees well with the values
reported by Lv \textit{et al.} \cite{Lv2008} and is about $8 \%$ different
than that reported by Faraoun \textit{et al.} \cite{Faraoun2006}.
For shear modulus, $G$, Poisson's ratio, $\nu$, and elastic modulus, $E$,
(Tabel \ref{tab:4}) no data is available for comparison. 

\begin{table}[h!]
\caption{Polycrystalline elastic moduli of $\eta$-Fe$_2$C (GPa).}
\begin{center}
\begin{tabular}{ccccc}
\hline
Source & $B \, (\text{GPa})$ & $G \, (\text{GPa})$ & $E \, (\text{GPa})$ & $\nu$\\
\hline
This study & 223 & 147 & 362 & 0.23 \\
\cite{Faraoun2006} & 243 & - & - & - \\
\cite{Lv2008} & 226 & - & - & - \\
\hline
\end{tabular} \\
\end{center}
\label{tab:4}
\end{table}
\subsection*{Nanoindentation}
\label{subsec:nanoires}
The average elastic modulus of martensite, determined from nanoindentation
tests carried out on the oil quenched samples was $E=203 \, \text{GPa}$
and it was used as the reference value in the subsequent calculations
of the volume fraction of carbides.
The average elastic modulus of retained austenite was $E=175 \, \text{GPa}$.
Similar values were reported for the elastic modulus of Fe-C austenite
from molecular dynamics calculations \cite{Oila2009}. 

\begin{table*}[ht!]
\caption{Retained austenite measurements (\%).}
\begin{center}
\begin{tabular}{cccccccccc}
\hline
Sample & DCT & SCT & Oil quenched \\
\hline
Retained austenite (\%) & $7.7 \pm 0.6$ & $9.3 \pm 0.7$ & $23.3 \pm 4.1$  \\
\hline
\end{tabular} \\
\end{center}
\label{tab:5}
\end{table*}

Considering the percentages of retained austenite determined by XRD (Table \ref{tab:5})
and the elastic modulus of each phase which contributes to the measured elastic modulus,
the volume fraction of carbides can be estimated using the rule of mixtures (Eq. \ref{eq:13}).
The phases considered are martensite ($E=203 \, \text{GPa}$),
retained austenite ($E=175 \, \text{GPa}$) and $\eta$-Fe$_2$C ($E=362 \, \text{GPa}$).
The resulting volume fractions of carbides calculated using the rule of mixtures
(Eq. \ref{eq:13}) are $f_{DCT}=0.20$ and $f_{SCT}=0.04$.

\begin{equation}
 E = \sum_{k=1}^{n}E_k \cdot v_k
\label{eq:13}
\end{equation}
\begin{tabbing}
 where:\= \\
 \> $E$ is the elastic modulus of composite\\
 \> $E_k$ is the elastic modulus of phase $k$\\
 \> $v_k$ is the volume fraction of phase $k$
\end{tabbing}

\begin{figure}[h!]
  \includegraphics[height=0.45\textwidth, angle=-90]{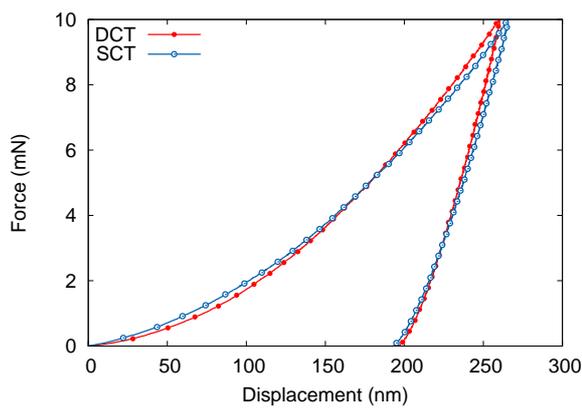}
  \caption{Load-displacement curves for the two steels.}
  \label{fig:Fig6}
\end{figure}

\begin{figure*}[ht!]
  \begin{center}
    \subfigure[]{\label{subfig:Fig7a}\includegraphics[width=0.45\textwidth]{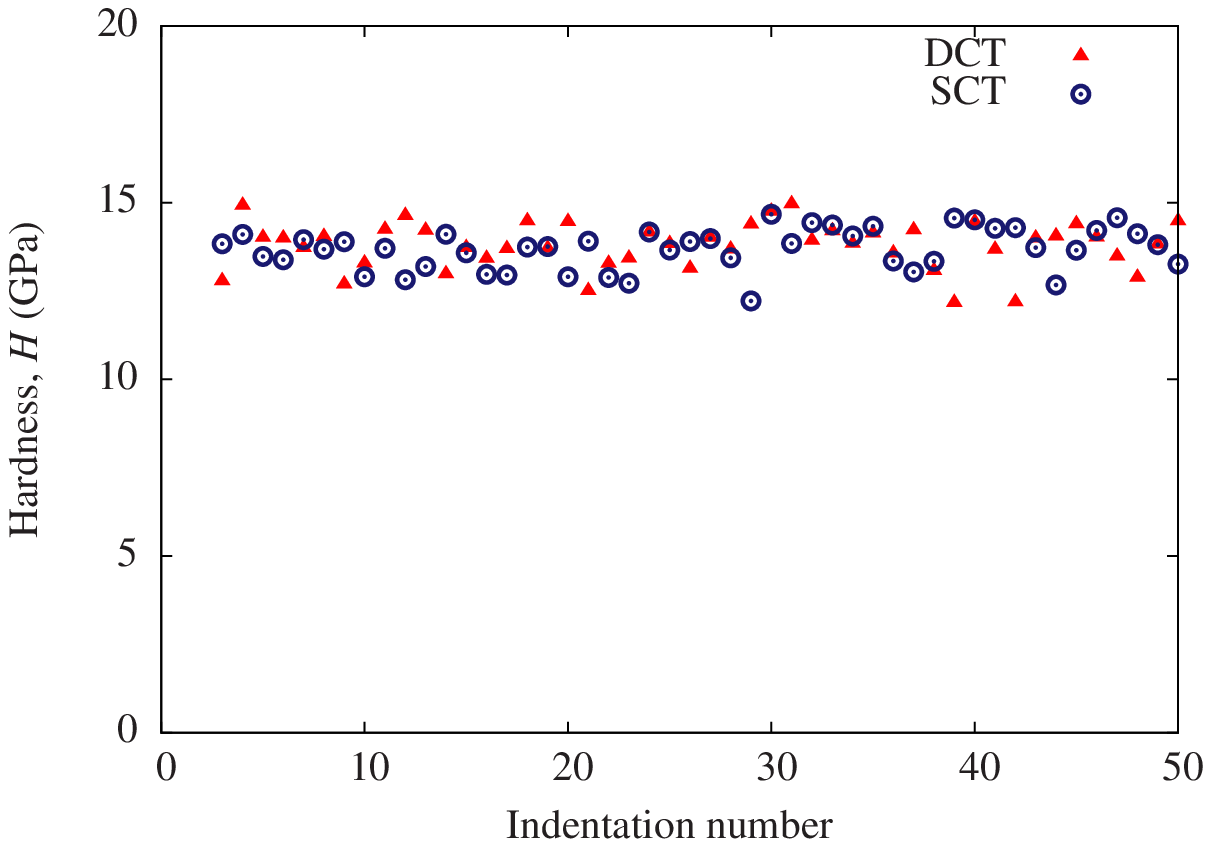}}
    \subfigure[]{\label{subfig:Fig7b}\includegraphics[width=0.45\textwidth]{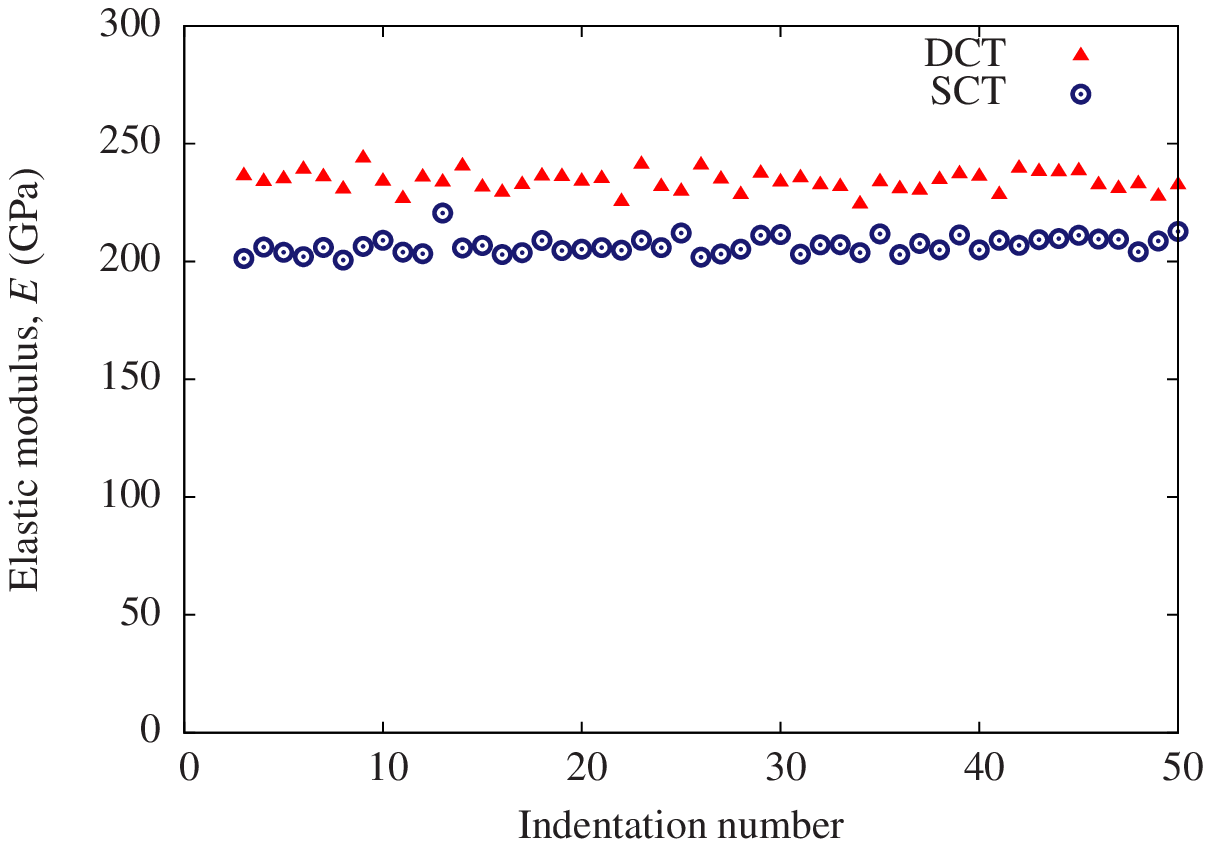}}
  \end{center}
  \caption{Nanoindentation results.
 (a) Hardness; (b) Elastic modulus.}
  \label{fig:Fig7}
\end{figure*}

\begin{figure*}[ht!]
  \begin{center}
    \subfigure[]{\label{subfig:Fig8a}\includegraphics[width=0.35\textwidth]{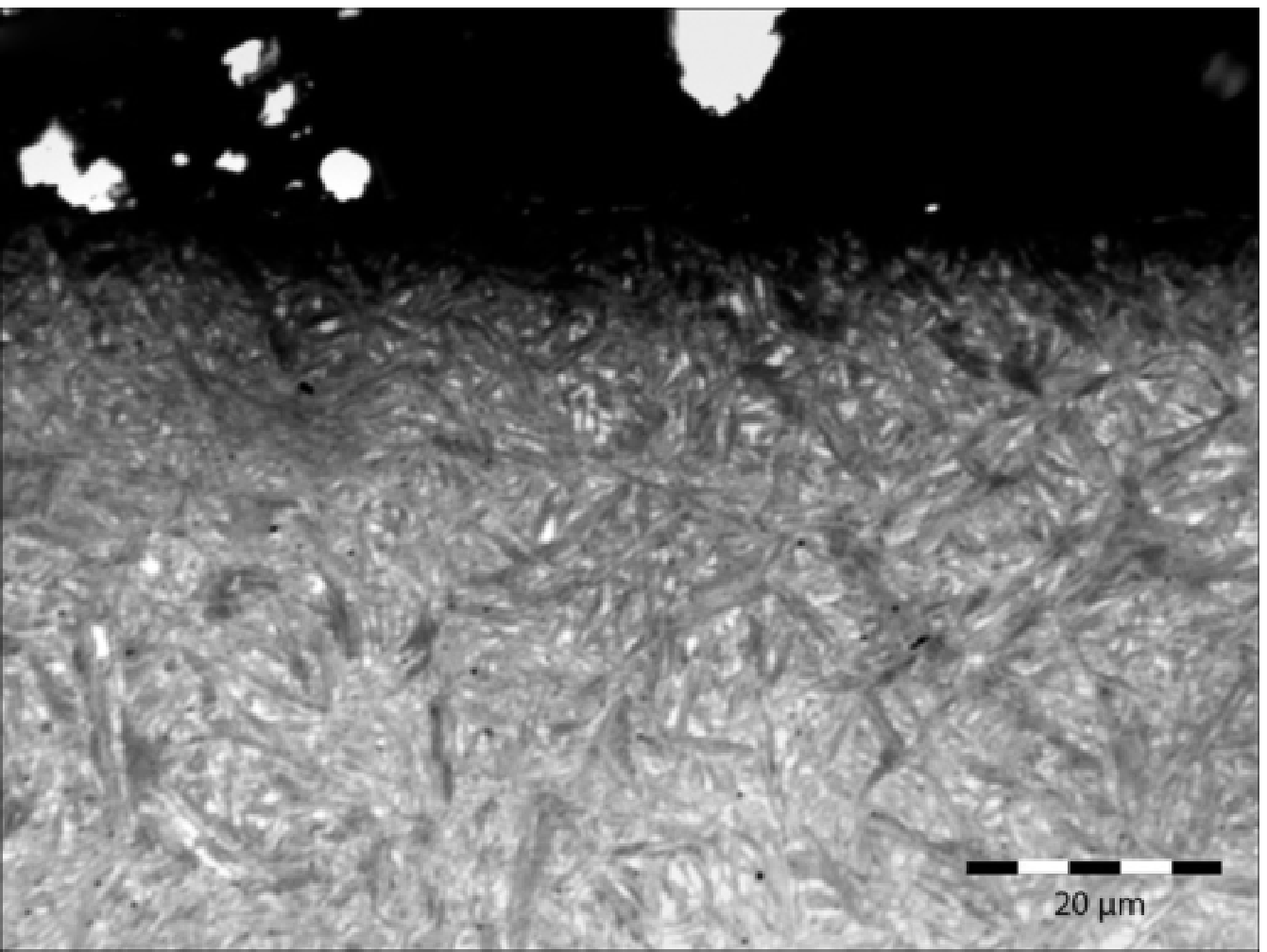}}
    \subfigure[]{\label{subfig:Fig8b}\includegraphics[width=0.35\textwidth]{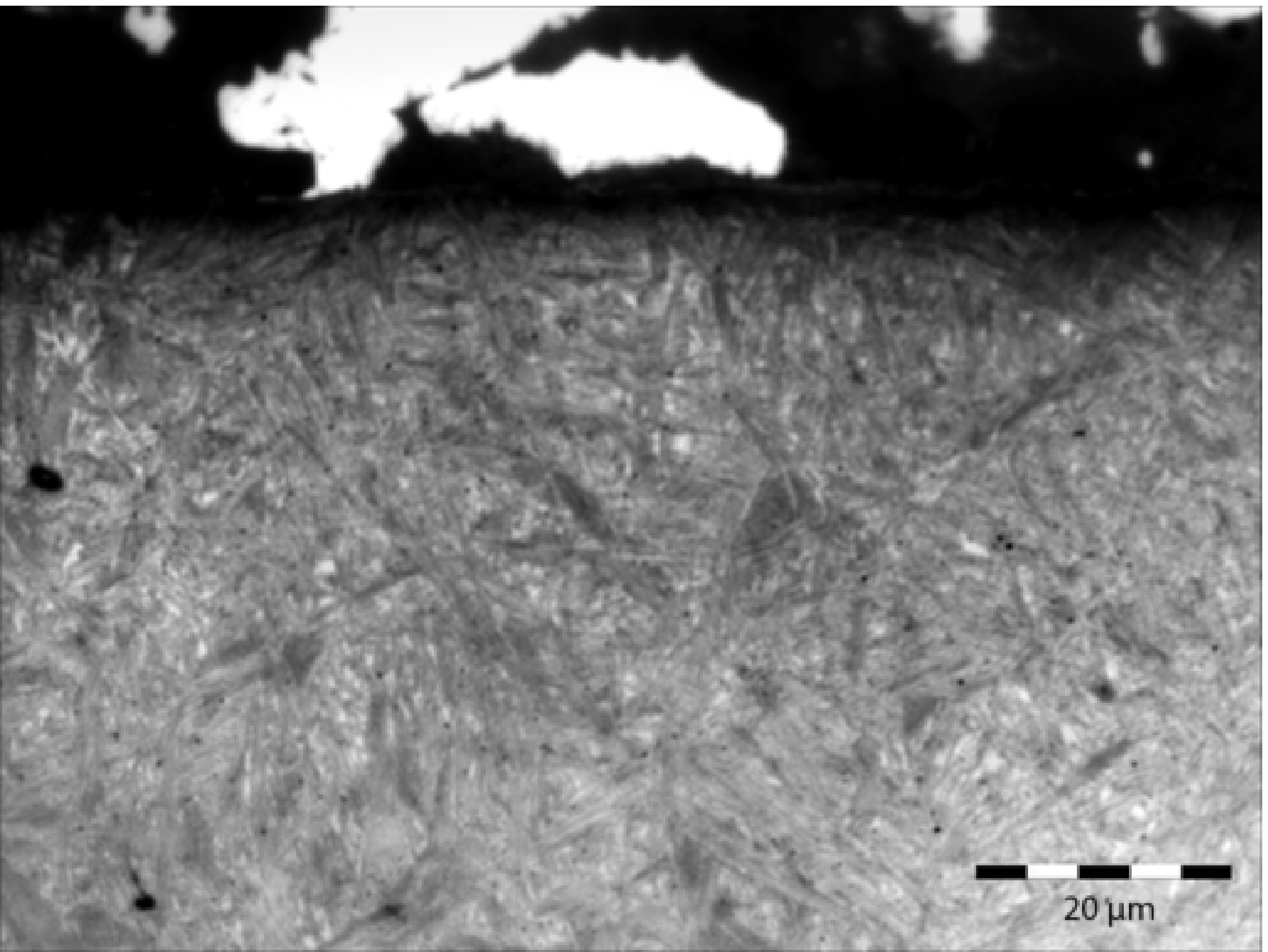}}
  \end{center}
  \caption{Microstructure of (a) SCT specimen; (b) DCT specimen. Picral etch.}
  \label{fig:Fig8}
\end{figure*}

These results show that only a small
amount ($4 \%$) of $\eta$-Fe$_2$C precipitates during SCT while during DCT
a large number of carbides will form ($20 \%$).
Typical load-displacement curves obtained for the SCT and DCT steels
are shown in Fig. \ref{fig:Fig6}. It can be seen that the total
displacement at the maximum applied load and the elastic recovery
are slightly larger for the SCT steel.
The measured hardness and elastic modulus are plotted in Fig. \ref{subfig:Fig7a}
and \ref{subfig:Fig7b}, respectively.
Both, hardness and elastic modulus show little scattering which indicate
that the microstructure (see Fig. \ref{fig:Fig8}) is relatively homogeneous.
The average hardness values are $H_{DCT}=13.8 \pm 0.6 \, \text{GPa}$ for DCT steel and
$H_{SCT}=13.7 \pm 0.6 \, \text{GPa}$ for SCT steel.
The average elastic modulus was $E_{DCT}=233.7 \pm 4.4 \, \text{GPa}$
for DCT steel and $E_{SCT}=206.7 \pm 3.7 \, \text{GPa}$ for SCT steel.

\subsection*{Surface contact fatigue}
\label{subsec:fatigueres}
\begin{figure*}[ht!]
  \begin{center}
    \subfigure[]{\label{subfig:Fig9a}\includegraphics[width=0.3\textwidth]{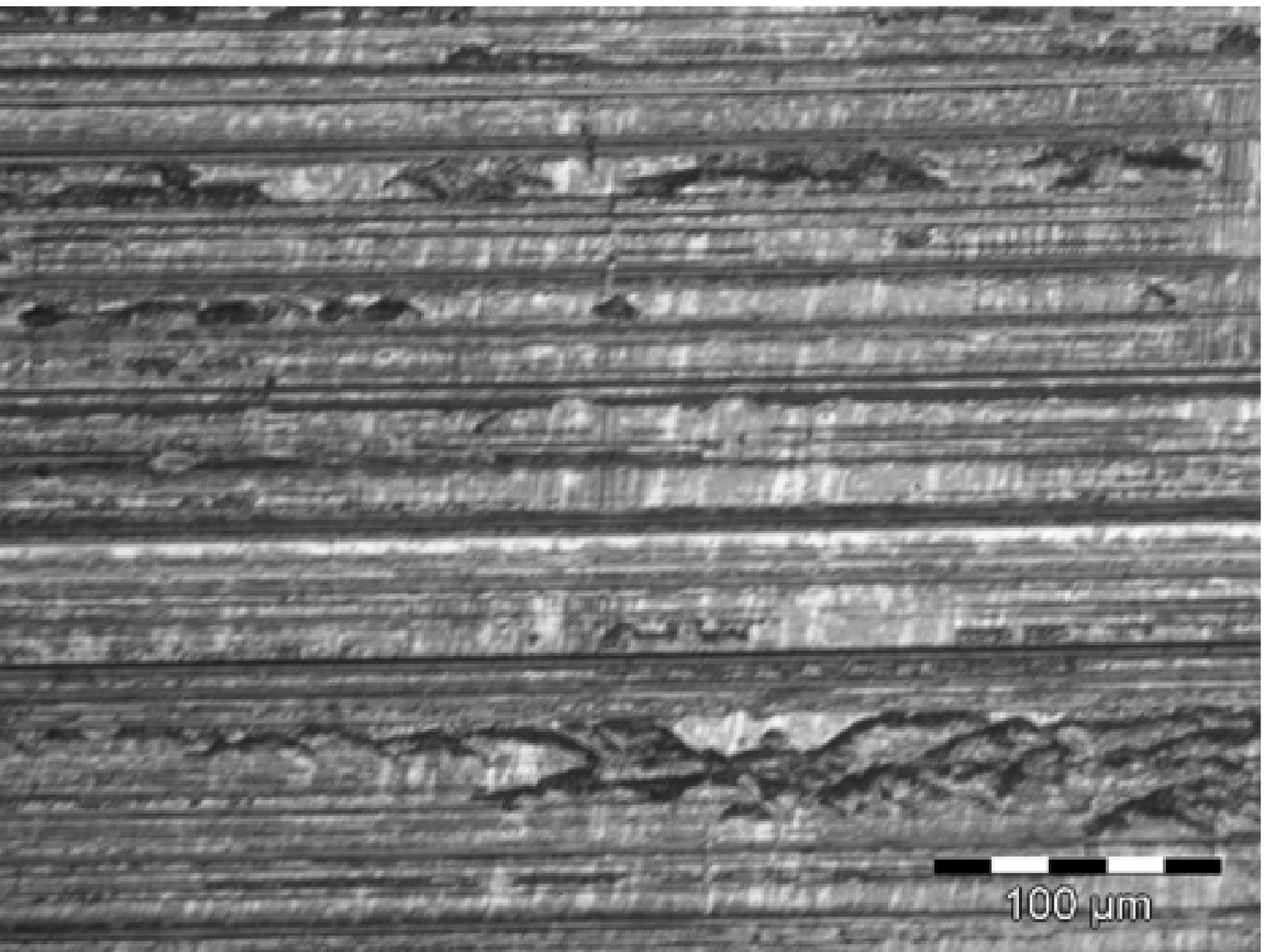}}
    \subfigure[]{\label{subfig:Fig9b}\includegraphics[width=0.3\textwidth]{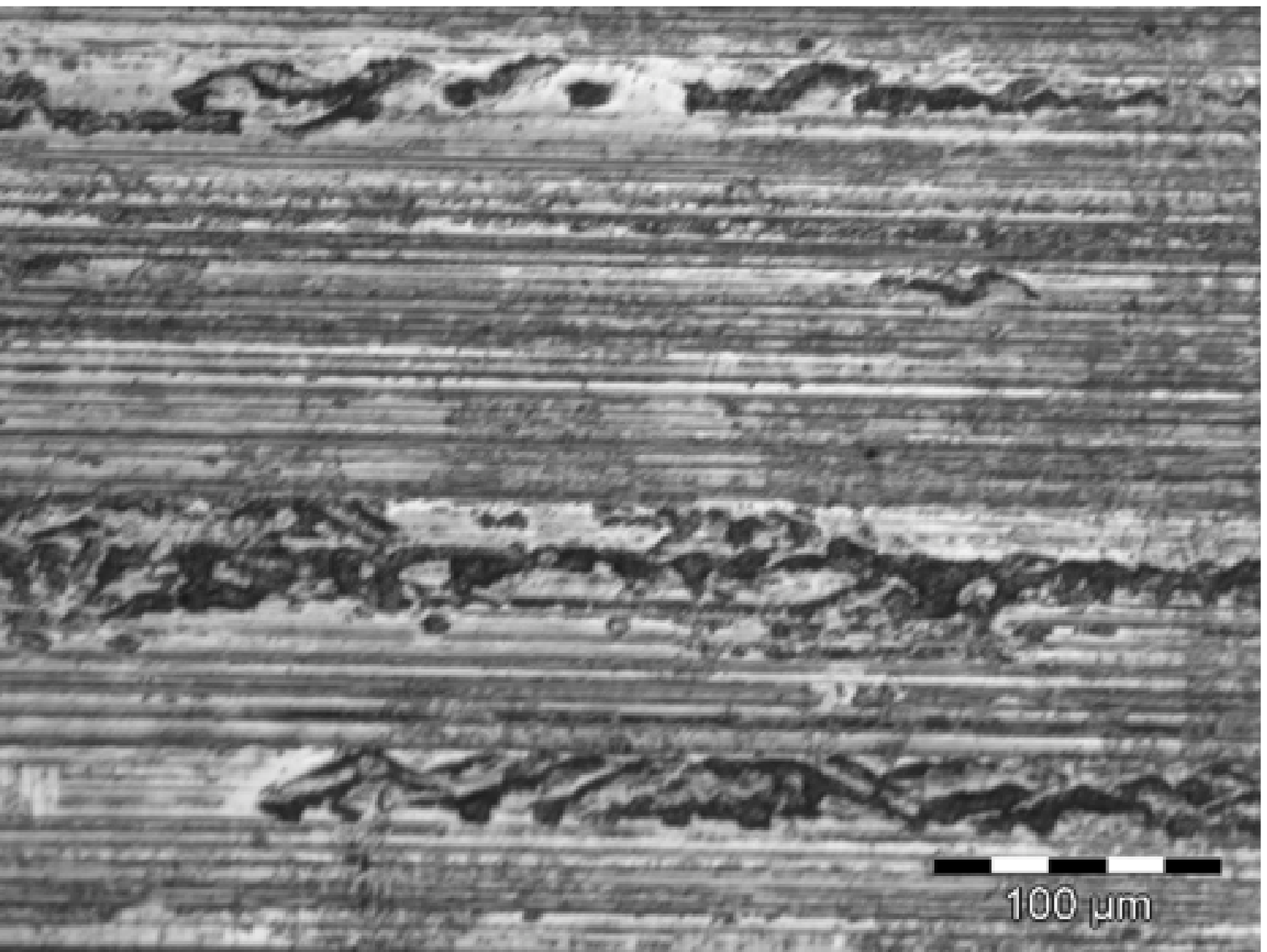}}
    \subfigure[]{\label{subfig:Fig9c}\includegraphics[width=0.3\textwidth]{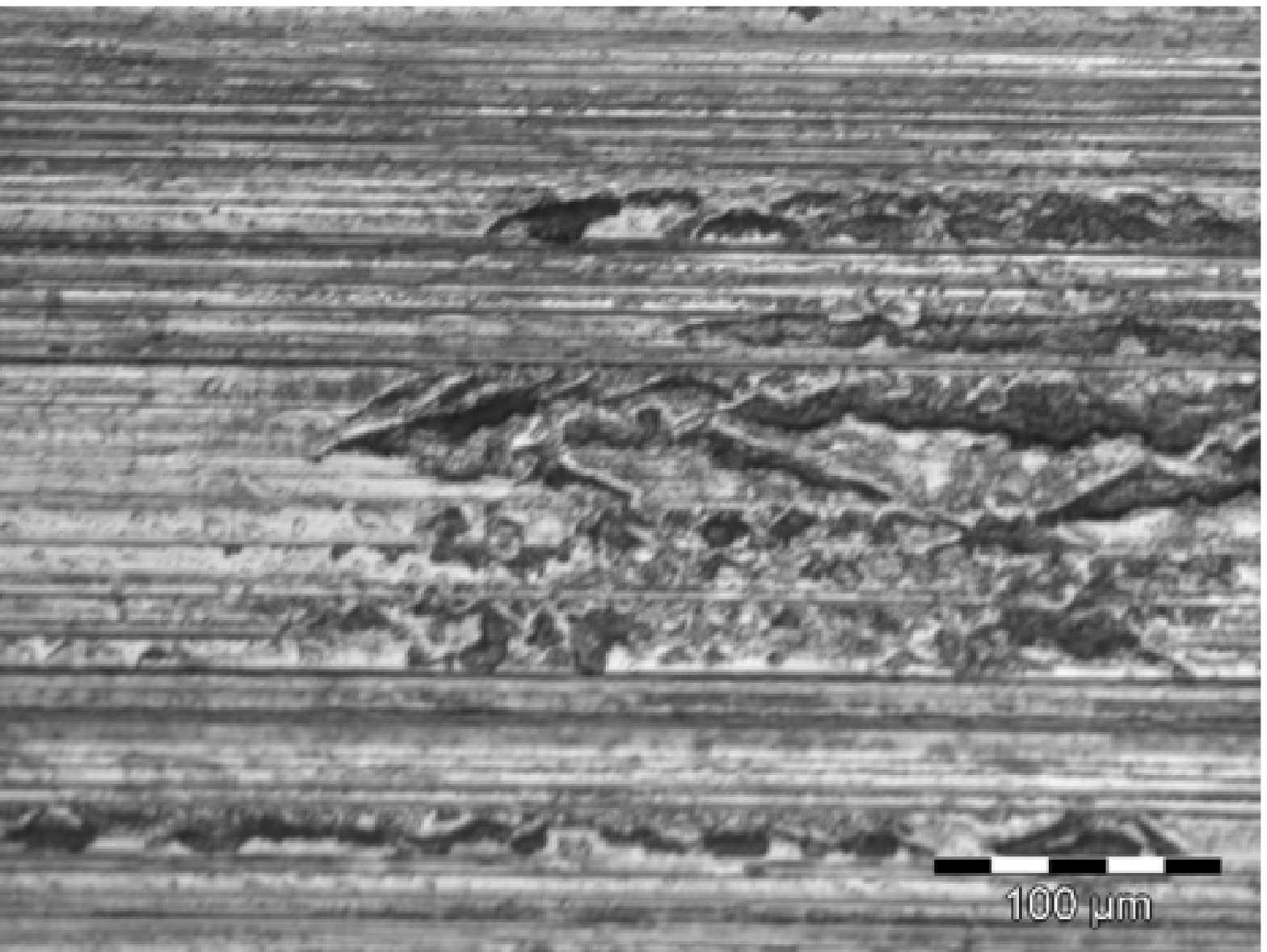}}
  \end{center}
  \caption{Light microscopy images showing the surface of (a) DCT sample;
   (b) SCT sample; (c) oil quenched sample.}
  \label{fig:Fig9}
\end{figure*}

Compared to conventional oil quenching, both cryogenic treatments lead to
a reduction of micropitting (see Fig. \ref{fig:Fig9}).
Table \ref{tab:6} shows the average area of micropitting measured for each specimen.   

Both cryogenic treatments are effective in improving the contact
fatigue resistance but due to different effects. 
Since the precipitation of $\eta$-Fe$_2$C in the SCT steel is not significant
($4 \%$) the improvement is probably due the transformation of retained austenite.
On the other hand, the intense precipitation of $\eta$-Fe$_2$C in the DCT steel
($20 \%$) increases the fracture toughness of martensite by a mechanism specific to
metal matrix composites: (1) crack deflection by the stiffer nanoparticle,
(2) crack trapping by nanoparticle which results in significant reduction
of stresses in the matrix and, (3) crack bridging ahead of the main crack tip. 

\begin{table}[h!]
\caption{Average micropitting area, $M \, (\%)$.}
\begin{center}
\begin{tabular}{cccc}
\hline
Sample & DCT & SCT & Oil quenched\\
\hline
$M \, (\%)$ & $4.2 \pm 1.2$ & $5.8 \pm 1.4$ & $8.2 \pm 1.5$ \\
\hline
\end{tabular} \\
\end{center}
\label{tab:6}
\end{table}

\section*{Conclusions}

In this work the elastic properties of $\eta$-Fe$_2$C have been determined
from \textit{ab initio} calculations. The hardness and elastic modulus
of a case carburised gear steel subjected to cryogenic treatments (SCT and DCT)
have been determined by nanoindentation.
Based on the elastic modulus of $\eta$-Fe$_2$C
derived from first principles the volume fraction of carbides was
estimated. It was found that the microstructure of the SCT steel contains
only $4\%$ of $\eta$-Fe$_2$C the microstructure of the DCT steel contains
$20\%$ of $\eta$-Fe$_2$C.
The precipitation of eta carbide in the DCT steel results in an increase
in elastic modulus but there is no difference in the hardness of
the DCT steel and SCT steel.

The micropitting tests carried out under EHL conditions showed
that cryogenic treatments improved the surface contact fatigue behaviour
of S156 case carburised steel. The average micropitting area was $8.2 \%$
for the oil quenched steel, $5.8 \%$ for the SCT steel and $4.2 \%$
for the DCT steel.
Both cryogenic treatments are effective in reducing micropitting
but the mechamisms involved are probably different.
The improved contact fatigue performance of the SCT steel
is due to the transformation of retained austenite while
in the DCT steel this is due to an increase in fracture toughness
as a result of eta carbide precipitation.
The nano-carbides act as reinforcements in the martensite matrix
by one of the mechanisms specific to composite materials:
(1) crack deflection, (2) crack trapping and, (3) crack bridging.  
\begin{acknowledgements}
The authors thank Frozen Solid for carrying out the cryogenic treatments.
Special thanks to Chris Aylott from the Design Unit - Newcastle University
for fruitful discussions and for carrying out the retained austenite measurements.
\end{acknowledgements}
\bibliographystyle{spphys}
\bibliography{AOJMaterSci13.bib}
\end{document}